\documentclass[aps,pra,twocolumn,showpacs,preprintnumbers,amsmath,amssymb]{revtex4-1}

\usepackage{hyperref}
\usepackage{graphicx}
\usepackage{graphics}
\usepackage{amsmath}

\begin{document}

\title{Local gravity measurement with the combination of atom interferometry and
Bloch oscillations}

\author{Ren\'{e}e Charri\`{e}re}
\author{Malo Cadoret}
\altaffiliation{Present address: Laboratoire Commun de M\'{e}trologie LNE-CNAM
61 rue du Landy, 93210 La plaine Saint Denis, France}
\author{Nassim Zahzam}
\author{Yannick Bidel}
\email{yannick.bidel@onera.fr}
\author{Alexandre Bresson}
\affiliation{ONERA, BP 80100, 91123 Palaiseau Cedex, France}

\begin{abstract}
We present a local measurement of gravity combining Bloch oscillations and atom
interferometry. With a falling distance of 0.8 mm, we achieve a sensitivity of
$2 \times 10^{-7}\;\text{g}$ with an integration time of 300 s. No bias
associated with the Bloch oscillations has been measured. A contrast decay with
Bloch oscillations has been observed and attributed to the spatial quality of
the laser beams. A simple experimental configuration has been adopted where a
single retro-reflected laser beam is performing atoms launch, stimulated Raman
transitions and Bloch oscillations. The combination of Bloch oscillations and
atom interferometry can thus be realized with an apparatus no more complex than
a standard atomic gravimeter.
\end{abstract}

\maketitle

\section{Introduction}

Gravimeters based on atom interferometry have shown impressive results with
record sensitivity ($\sim 10^{-8}\;\text{g}\,\text{Hz}^{-1/2}$) \cite{best sens
muller,limit sens gravi syrte} and accuracy ($\sim 10^{-9}\;\text{g}$)
\cite{gravi peter,comp gravi syrte}. However, a falling distance of at least
$\sim 7\; \text{cm}$ is required to reach those performances preventing them
from being miniaturized and making local gravity measurement. Atomic
gravimeters based on Bloch oscillations \cite{mesure g biraben,mesure g tino}
can measure gravity with an interaction distance of few micrometers but the
performances are reduced compared to gravimeters based on atom interferometry.
The best Bloch oscillations gravimeter reaches an accuracy of $\sim
10^{-7}\;\text{g}$ and a sensitivity of $\sim 10^{-7}\;\text{g}$ with a
measurement time of one hour \cite{mesure g tino}.

In this article, we present an atom gravimeter combining atom interferometry
and Bloch oscillations. This scheme allows to associate the sensitivity
provided by atom interferometry and the locality provided by Bloch
oscillations.

The combination of Bloch oscillations and atom interferometry has already been
demonstrated in order to obtain large momentum transfer atomic beam splitter
\cite{Bloch splitter biraben,Bloch splitter chu}. The use of optical lattices
as waveguide and beam splitter for atom interferometry has also been studied
theoretically \cite{theorie latt interf}. In those works, the two arms of the
atomic interferometer are guided by different lattices in which Bloch
oscillations occur. The combination of Bloch oscillations and atom
interferometry has also been demonstrated in the frame of the fine structure
constant measurement \cite{det st fine biraben}. In this experiment,  a
Ramsey-Bord\'{e} interferometer is used. It is characterized by the fact that
the two arms are in the same quantum state in the middle of the interferometer.
Thus, only one lattice is needed to guide the atoms in the central part of the
interferometer.

In this article, we will adapt this previous combination of Bloch oscillations
and Ramsey-Bord\'e atom interferometer for gravity measurements. In the first
part, we will present the configuration which we have chosen to combine atom
interferometry and Bloch oscillations. We will provide also the expression of
the phase of the interferometer. Our experimental set-up will be described in
the second part. Then, we will present the gravity measurements. Finally, we
will analyze the impact of Bloch oscillations on the contrast of the
interference fringes and on the bias concerning gravity measurements.

\section{Gravimeter scheme combining atom interferometry and Bloch oscillations}

\begin{figure}[h]
 \includegraphics[scale=0.65]{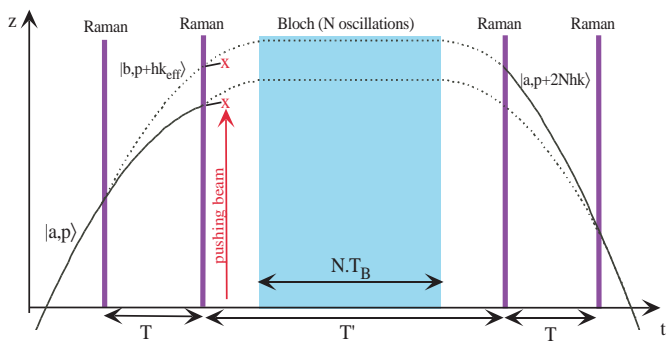}
 \caption{\label{interf}Gravimeter scheme combining Ramsey-Bord\'{e} interferometer and Bloch oscillations.}
\end{figure}

We use a Ramsey-Bord\'{e} atom interferometer \cite{ramsey borde} composed of
four $\pi/2$ Raman laser pulses separated by the times T, T' and T (see figure
\ref{interf}). The Raman laser drives a transition between the internal state
$|a\rangle$ with a momentum $p$ and the internal state $|b\rangle$ with a
momentum $p+\hbar k_{\mathrm{eff}}$ where $k_{\mathrm{eff}}$ is the effective
wave vector associated with the Raman laser. A pushing beam at resonance with
the atoms in the state $|a\rangle$ is applied in the middle of the
interferometer. After this step, only atoms in the state $|b\rangle$ remain in
the interferometer. If the two photons detuning of the Raman laser is small
compare to the Rabi frequency, one can show that the probability to be in the
state $|a\rangle$ at the end of the interferometer is :
\begin{equation}
P_a=\frac{1}{2}+\frac{1}{4}\cos(\Phi)
\end{equation}
where the phase of the interferometer $\Phi$ is equal to :
\begin{equation}
\Phi=(k_{\mathrm{eff}} g - \alpha)\, T\, (T+T')
\end{equation}
where $\alpha$ is the radio frequency chirp applied to the Raman laser which
compensates the Doppler effect due to gravity acceleration $g$. In this
expression, the sign conventions are: $k_{\mathrm{eff}}>0$ when
$k_{\mathrm{eff}}$ is downward and $g>0$.

In order to gain sensitivity without increasing the falling distance, the atoms
are placed in a laser standing wave between the two pairs of $\pi/2$ pulses.
The standing wave is produced by two counterpropagating laser beams. If one
neglects spontaneous emission, the atoms interact with the process of photon
absorption from one beam and stimulated emission into the other beam. The
result is a change of momentum of the atoms of $2\hbar k$ where $k$ is the wave
vector of the laser. This process is resonant when the velocity of the atoms
relative to the lattice is equal to plus or minus the recoil velocity
$v_{rec}=\hbar k/m$. Under gravity, the velocity of the atoms decreases until
$-v_{rec}$ when the atoms become in resonance with the two photons transition.
At this point, an adiabatic transition occurs and the velocity of the atoms
increases by $2 v_{rec}$. This process repeats itself every $T_B=2 v_{rec}/g$
which is called the Bloch period. This phenomenon where the atom velocity
oscillates between $-v_{rec}$ and $v_{rec}$ by exchanging photons every $T_B$
with the laser is called Bloch oscillation. This process can be very efficient,
20 000 Bloch oscillations have been experimentally demonstrated \cite{Bloch
oscillation nagerl}. Moreover, the variation of atom velocity which occurs
after $N$ Bloch oscillations is perfectly known and is equal to $2Nv_{rec}$.

In the middle of the interferometer, the atoms interact with the Bloch laser
beams during a time $N\,T_B$. After this interaction, the velocity change
compared to free falling atoms is equal to $2N\,v_{rec}$ leading to an
additional phase shift on the interferometer of $-k_{\mathrm{eff}} 2 N v_{rec}
T$. A frequency jump of $\Delta \omega \simeq - k_{\mathrm{eff}} 2 N v_{rec}$
is applied to the Raman frequency in order to keep the Raman laser at resonance
with the atoms after the change of velocity produced by the Bloch oscillations.
Finally, one obtains for the phase of the interferometer including Bloch
oscillations :
\begin{equation}\label{phase interf}
\phi=(k_{\mathrm{eff}} g - \alpha) T (T+T')-(k_{\mathrm{eff}} 2 N v_{rec}+\Delta \omega)T
\end{equation}
The additional phase shift caused by Bloch oscillations is well known because
it depends on the laser wavelength and can be canceled experimentally by
applying the appropriate frequency jump on the Raman frequency.

In the configuration in which the atoms are launched and then oscillate in a
fixed optical lattice (see figure \ref{interf}), one can increase the scaling
factor of the interferometer by increasing $T'$ without increasing the falling
distance.  It is therefore possible to have a gravimeter with a scaling factor
not limited by the vertical size of the instrument.

\section{Experimental apparatus}

In this part, the whole experimental apparatus is described: the laser source
used to manipulate rubidium 87 atoms, the source of cold atom and finally the
experimental procedure which combines atom interferometer and Bloch
oscillations.

\subsection{Laser sources}
The laser for trapping, cooling, detecting and pushing the atoms is realized
with the frequency doubled fiber bench at 1560 nm described in reference
\cite{laser onera}.

\begin{figure}[h]
 \includegraphics[scale=0.67]{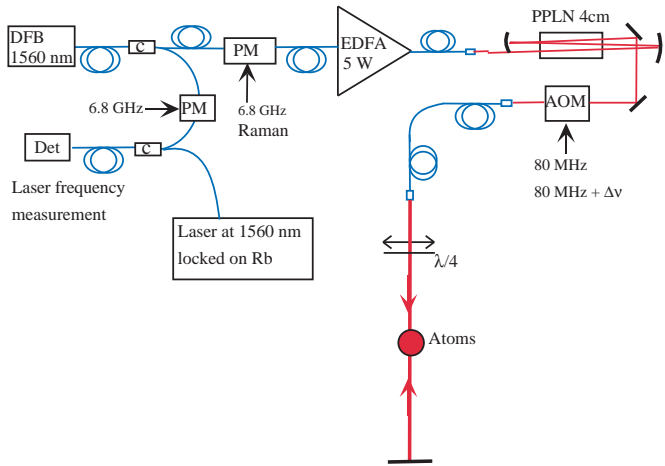}
 \caption{\label{laserbloch} Scheme of the Raman/Bloch laser (PM: phase modulator, Det: fast photodiode,
  DFB: DFB laser diode, EDFA: erbium doped fiber amplifier, AOM: acousto-optic modulator,
  PPLN: periodically poled lithium niobate crystal, c: fiber coupler).}
\end{figure}

The Raman and Bloch laser are coming out from the same laser source and are
also obtained by frequency doubling a fiber bench laser at 1560 nm (see figure
\ref{laserbloch}). A distributed feedback (DFB) laser diode at 1560 nm is
amplified in a 5 W erbium doped fiber amplifier (EDFA). Then, the amplified
laser is frequency doubled in a periodically poled lithium niobate crystal
(PPLN) used in a triple pass geometry. The laser is directed through an
acousto-optic modulator (AOM) which controls the intensity of the laser light.
The laser beam interacts with the atoms with a waist of 4.5 mm, a circular
polarization and a maximum power of 0.2 W. The beam has been collimated with a
doublet corrected from abberations. The laser beam is retro-reflected by a
mirror placed on a vibration isolation table.

When one wants to address a stimulated Raman transition between the states F=1
and F=2, a radio frequency wave at 6.8 GHz is sent to a fiber phase modulator
placed after the DFB laser diode. This modulation creates sidebands at 6.8 GHz
on the spectrum of the laser. The radio frequency power is set so that the
power ratio at 780 nm between the first sidebands and the central peak is 0.5.

The measurement of the Bloch/Raman laser wavelength is obtained by mixing it
with a reference laser at 1560 nm locked on a rubidium transition and by
acquiring the beat note with a fast photodiode. The frequency of the beat note
is then measured with a 7 GHz spectrum analyzer. In order to measure detuning
bigger than 7 GHz, a fiber phase modulator fed with a 6.8 GHz radio frequency
wave is inserted on the Raman/Bloch laser. Then, the frequency of the beat note
between the reference laser and one of the sidebands of the Bloch/Raman laser
is measured. The frequency of the Bloch laser is then given by
$\nu_{ref}+2\nu_{beat}+2n\times 6.8\,\text{GHz}+\nu_{AOM}$ where $\nu_{ref}$ is
the frequency of the reference laser at 780 nm, $\nu_{beat}$ is the frequency
of the beat note, $n$ is the order of the sideband which participates to the
beat note and $\nu_{AOM}$ is the frequency of the acousto-optic modulator. With
this technique, one can measure the absolute frequency of the Bloch/Raman laser
over a large range of detuning (0-60 GHz) and with an accuracy of about 1 MHz,
limited by the frequency uncertainty of the reference laser. In the results
presented in this article, the Bloch/Raman laser is blue-detuned by 58 GHz from
the F=2 $\rightarrow$ F'=2 transition.

\subsection{Source of cold atoms and launch with Bloch oscillations}

The source of cold atoms is a magneto-optical trap of rubidium 87 loaded from a
background vapor. The atoms are then further cooled down in an optical molasses
to a measured temperature of $1\;\mu \text{K}$. At the end of the optical
molasses the atoms are in all the sublevels of the state F=2. A vertical
magnetic field of quantification of 60 mG is then applied.

\begin{figure}[h]
 \includegraphics[scale=0.6]{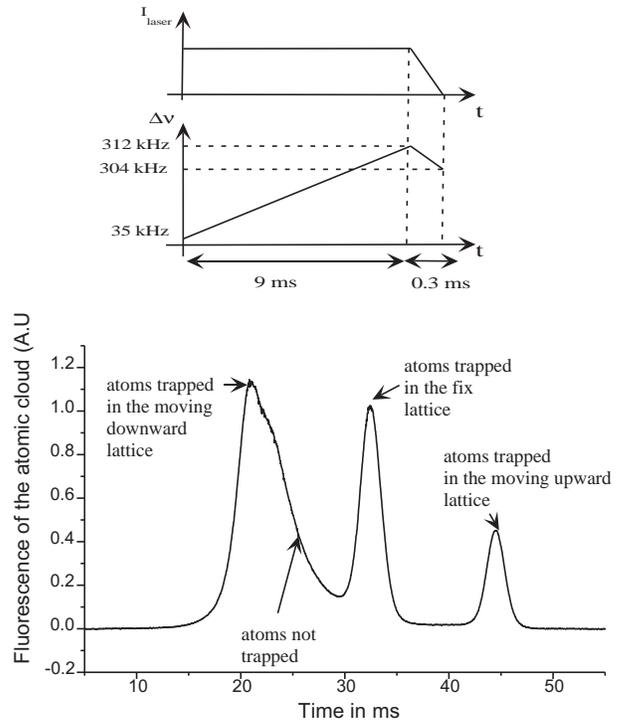}
 \caption{\label{launch} Launch of the atoms with the Bloch laser. Up: Temporal characteristics of the moving optical lattice during the launch.
 Down: Fluorescence of the atomic cloud passing through a 0.5 mm light sheet situated 32 cm below the magneto-optical trap.
 }
\end{figure}

The launch of the atoms is realized with the Bloch laser. For this, two
frequencies separated by $\Delta \nu$ are applied to the acousto-optic
modulator. In this case, the Bloch laser is composed mainly of two frequencies
separated by $\Delta \nu$. Thus, the retro-reflected Bloch laser is composed of
three standing waves: one fix, one traveling upward at a velocity of
$\lambda\,\Delta \nu/2$ and one traveling downward with the opposite velocity.
Depending on their velocity, the atoms which interact with this modulated
lattice will be trapped in the fix, the moving upward or moving downward
lattice. In order to realize the launch of the atoms, the modulated lattice is
applied just after the molasses during 9 ms. The frequency $\Delta\nu$ is
ramped from 35 kHz to 312 kHz in 9 ms. After, the lattice is turn off
adiabatically in 0.3 ms. The time of flight signal obtained after this launch
procedure is represented on figure \ref{launch}. We observe clearly three peaks
corresponding to atoms trapped in the three different lattices. With this
technique, $9.4\%$ of the atoms are launched at a velocity of 0.12 m/s. The
velocity width of the launched atoms is equal to $2 v_{rec}$ given by the width
of the first band of the lattice. For the used lattice depth, only atoms from
the first band are trapped.

\subsection{Ramsey-Bord\'{e} interferometer and Bloch oscillations}

After the launch, we apply two $\pi/2$ Raman pulses which drive a Raman
transition from the state $|a\rangle=|F=2,m_F=0\rangle$ to the state
$|b\rangle=|F=1,m_F=0\rangle$. The pulses are separated by the time T and have
a duration of $30\; \mu \text{s}$. Then, we send a vertical pushing beam which
blows the atoms remaining in the $F=2$ state and thus selects the atoms in the
state $|F=1,m_F=0\rangle$.

When the atoms reach the top of their trajectory, the Bloch laser is turned on
adiabatically in $200\;\mu \text{s}$. The power of the Bloch laser is set to a
point corresponding to a trade off between the losses coming from spontaneous
emission which increase with the power and the Landau-Zener losses \cite{Bloch
Salomon PRA} which decrease with the power. This corresponds to a lattice depth
$U_0$ of approximately two recoil energy $U_0 \sim 2 E_{rec}$. For this depth,
the atoms are only trapped in the first band of the lattice. The atoms in the
upper band are not trapped because Landau-Zener losses are too big. The lattice
is turned on for a duration of $N T_B$ during which $N$ Bloch oscillations
occur. Our experimental parameters give $T_B=1.20\;\text{ms}$. The Bloch laser
is then turned off adiabatically in $200\; \mu\text{s}$. Finally, two other
$\pi/2$ Raman pulses of duration $30\;\mu\text{s}$ separated by a time T are
applied. The total duration of our interferometer is equal to
$25\,\text{ms}+N\,T_B$. During this time, the atoms are moving over a distance
of $0.8 \;\text{mm}$.

Finally, we measure the proportion of atoms in the state F=2 and F=1. This step
is performed with three pulses of a vertical retro-reflected beam of durations
of 2 ms, 0.1 ms and 2 ms. The first and the last pulses are resonant with the
F=2$\rightarrow$F'=3 transition and the middle pulse is resonant with the
F=1$\rightarrow$F'=2 transition and puts the atoms from the state F=1 to the
state F=2. The fluorescence of the atoms during the first pulse is proportional
to the number of atoms in the state F=2 and the fluorescence collected during
the last pulse is proportional to the total number of atoms.

\section{Gravity measurement}

\begin{figure}[h]
 \includegraphics[scale=0.92]{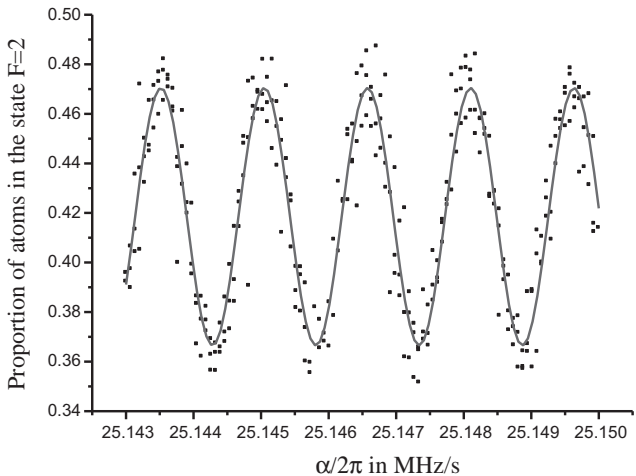}
 \caption{\label{frange} Proportion of atoms in the state F=2 versus the frequency chirp $\alpha$ applied to the Raman frequency
  (T=6 ms, T'=103 ms, 75 Bloch oscillations).
 The solid line is a fit with a sinus function. The value of the central fringe given by the fit is $\alpha_0=25 146 575 \pm 5\,\text{Hz}/\text{s}$.}
\end{figure}

The proportion of atoms in the state F=2 is measured for different values of
the radio frequency chirp $\alpha$ applied to the Raman frequency. Each
measurement takes 1 second corresponding to our experimental cycle time. A
typical signal is represented on figure \ref{frange}. The data are fitted with
the function $P_m+\frac{C}{2}cos((\alpha-\alpha_0)T(T+T'))$ where C is the
contrast of the fringes and $\alpha_0=k_{\mathrm{eff}} g$ is the value of the
central fringe giving a measurement of gravity. The parameters of the fit are
$P_m$, $C$ and $\alpha_0$. The error bars given in the different results
correspond to the error of the fit.

\begin{figure}[h]
 \includegraphics[scale=0.85]{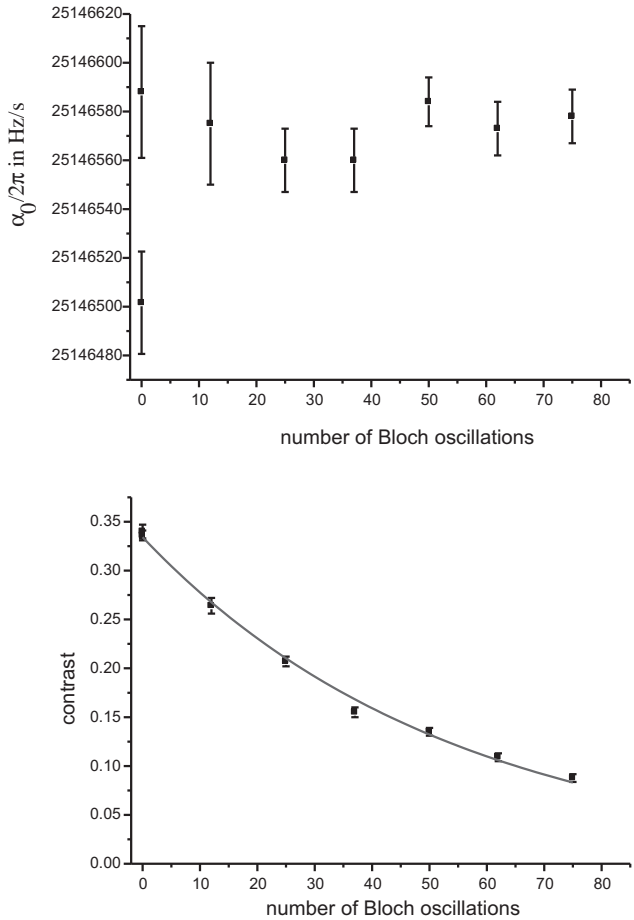}
 \caption{\label{alphaCversusN} Value of the central fringe $\alpha_0=k_{eff} g$ and contrast $C$ versus the number of Bloch oscillations ($T=6\,\text{ms}$,
  $T'=13\,\text{ms} + N \times 1.20\,\text{ms}$). Each value has been obtained with a fit over $\sim 150$ measurements.
 The solid line on the contrast graph is an exponential fit with a decay constant of $N_0=54$.}
\end{figure}

We have measured interference fringes for different number of Bloch
oscillations in the lattice. The measurement of the position of the central
fringe $\alpha_0$ and the contrast $C$ of the fringes are reported on figure
\ref{alphaCversusN}. One does not note any significant variation of $\alpha_0$
with the number of Bloch oscillations. We can presume that there is no bias
associated with the insertion of Bloch oscillations in the interferometer at
the level of precision in our experiment. This point will be discussed in
details in part V.

However, the contrast of the fringes decreases when the number of Bloch
oscillations increase. The decay is well fitted with an exponential function
which has a decay constant of $N_0=54$. This limitation prevents from
increasing significantly the sensitivity of the interferometer with Bloch
oscillations because the gain in scale factor is partially compensated by the
contrast drop. The best sensitivity (statistical uncertainty given by the fit)
obtained is $\sim 2\times 10^{-7}$ g in 300 s with 75 Bloch oscillations
(T'=103 ms) and T=6 ms. Our sensitivity is limited by our detection noise which
is equal to $\sigma_{P_2}=0.014$ (rms value of the noise on the proportion of
atoms in the state F=2). The implementation of a better detection system
\cite{det kasevich} should give a better sensitivity with an improvement of at
least 10.

\section{Bias associated with the Bloch oscillations}

The insertion of Bloch oscillations in the interferometer is responsible of
additional systematic effects compared to an atom interferometer alone. They
come from the uncertainty of the Raman frequency jump $\Delta\omega$ and from
the uncertainty of the velocity variation during the Bloch interaction. An
uncertainty on the velocity variation $\delta v$ gives an uncertainty on the
phase $\delta \phi=k_{\mathrm{eff}}\delta v\,T$ and thus an uncertainty on the
gravity measurement $\frac{\delta g}{g}=\frac{\delta v}{g(T+T')}$. In this
part, the different sources of error will be examined. All numerical results
will be given for 75 Bloch oscillations ($T'=103\,\text{ms}$) and
$T=6\,\text{ms}$.

\subsection{Uncertainty of the Raman frequency jump}

Using equation \ref{phase interf}, one obtains that an uncertainty $\delta
\omega$ on the Raman frequency jump gives an uncertainty of gravity measurement
equal to:
\begin{equation}
\frac{\delta g}{g}=-\frac{\delta\omega}{k_{\mathrm{eff}}g(T+T')}
\end{equation}
Our radio frequency chain performs the frequency jump with an uncertainty of
0.3 Hz, leading to an error on gravity measurement of $1.1\times 10^{-7}$ g.

\subsection{Uncertainty on the laser frequency}

The uncertainty on the laser frequency $\delta\nu_L$ leads to an uncertainty on
the recoil velocity and thus an uncertainty on the velocity variation during
the Bloch interaction $\delta v=\frac{\delta\nu_L}{\nu_L}2N\,v_{rec}$. One
obtains finally an uncertainty on the gravity measurement equal to :
\begin{equation}
\frac{\delta g}{g}=-\frac{\delta \nu_L}{\nu_L} \frac{N\,T_B}{(T+T')}
\end{equation}
In our experiment, the frequency of the Bloch laser is known with an
uncertainty of 10 MHz corresponding to the natural drift of the laser between
the laser frequency measurement (see III.A) and the gravity measurement. This
gives an error on the gravity measurement equal to $2.2\times 10^{-8}$ g. A
lock technique easily implementable on our setup can reduce the uncertainty on
the laser frequency down to 1 MHz and thus gives an error on gravity of
$2.2\times 10^{-9}$ g.

\subsection{Nonadiabaticity}

Landau-Zener losses and nonadiabaticity in turning off and on the lattice
produce atoms with a velocity $2(N+\Delta)\,v_{rec}$ where $\Delta$ is an
integer. These atoms have a phase shift equal to $k_{\mathrm{eff}}
2\Delta\,v_{rec} T$ which is large compare to $2\pi$ for usual time T used in
the experiment. This effect can thus give an error on the gravity measurement
but can be canceled by choosing a time T multiple of
$\frac{2\pi}{k_{\mathrm{eff}} 2\,v_{rec}}$. Experimentally, one did not see any
effect by choosing the appropriate T at the level of accuracy of our
experiment.

\subsection{Gouy phase}

The laser beam interacting with the atoms is not exactly a plane wave but a
gaussian beam. Thus, due to the Gouy phase, the velocity variation of atoms
after the process of absorption and stimulated emission will be not exactly $2
v_{rec}$. For atoms located in the waist of the laser beam, one obtains a
velocity variation of $2 v_{rec} (1-\frac{2}{k^2\,w^2})$ where $w$ is the waist
of the laser beam . This leads to a bias on the gravity measurement equal to:
\begin{equation}
\frac{\delta g}{g}=\frac{2N\,T_B}{k^2w^2(T+T')}
\end{equation}
For our experimental parameters, this error is equal to $1.3\times 10^{-9}$ g.

\subsection{Longitudinal dipolar electric force of the Bloch beam}

The vertical gradient of intensity of the Bloch Beam is responsible of a force
changing the velocity of the atoms during the Bloch interaction and thus
inducing a bias on the gravity measurement. For a gaussian beam, the intensity
varies longitudinally on the distance scale given by the Rayleigh distance
$Z_R$ which is in our case equal to 82 m. Thus, the order of magnitude of the
velocity variation after the Bloch interaction is equal to:
\begin{equation}
\delta v \lesssim \frac{U_0}{m\,Z_R} N\,T_B
\end{equation}
where $m$ is the atom mass. The error on gravity measurement is given by:
\begin{equation}
\frac{\delta g}{g} \lesssim \frac{U_0}{m\,Z_R\,g} \frac{N\,T_B}{T+T'}
\end{equation}
For our experimental parameters, one obtains $\delta g \lesssim 5.5\times
10^{-8}$ g. This error is not negligible and if one wants to reach the level of
accuracy of $10^{-9}$ g, the position of the atoms compared to the waist of the
Bloch laser should be known precisely in order to obtain a precise estimation
of the gradient of intensity. It is also possible to eliminate this systematic
effect by noticing that the dipolar force changes sign with the sign of the
laser detuning. Measuring gravity with a positive and a negative detuning of
the Bloch laser should allow to cancel the effect of the dipolar force.

\subsection{Index of refraction}

The rubidium gas coming from the vapor background and the cold atomic cloud
makes the Bloch beams to propagate in a medium with an index of refraction
leading to a slight modification of the recoil velocity. This problem is
studied in detail in the reference \cite{PRA h sur m}. For our experimental
parameters, the relative effect on the recoil velocity is approximately equal
to $1\times 10^{-9}$ which gives an uncertainty on the gravity measurement
equal to $1\times 10^{-9}$ g.

\subsection{Conclusion on the systematic effects}

The table \ref{table} gives a summary of the systematic effects coming from
Bloch oscillations implementation. The error budget gives a total error of
$1.2\times 10^{-7}$ g dominated by the Raman frequency jump. This point can be
solved easily with a better radio frequency chain. The other sources of
systematic effects show an error around few $10^{-8}$ g dominated by the
vertical gradient of intensity.

The measurement of gravity versus the number of Bloch oscillations which
indicates no significant dependency with the number of Bloch is compliant with
our error budget.

\begin{table}
\begin{tabular}{|l|c|}
  \hline
  source & error ($\times 10^{-9}$ g) \\
  \hline
  Raman frequency jump& 110\\
  Laser frequency & 22  \\
  Gouy Phase & 1\\
  Dipolar force & $\lesssim 55$\\
  Index of refraction &  1\\
  \hline
  \end{tabular}
\caption{Systematic effects coming from the Bloch oscillations on the gravity
measurement.}\label{table}
\end{table}

\section{Contrast decay associated with the Bloch oscillations}

The insertion of Bloch oscillations in the interferometer can induce a contrast
decay only via a random velocity variation $\delta v$ which gives a random
phase $k_{eff} \delta v T$. This random velocity variation can come from
spontaneous emission or from spatial defaults on the Bloch beams. One can show
that the contrast decay is given by the Fourier transform of the distribution
probability of velocity variation:
\begin{equation}
C=C_0 \int P(\delta v)cos(k_{eff}\delta v T)\,d\delta v
\end{equation}
where $C_0$ is the contrast without random velocity jump.

\begin{figure}[h]
 \includegraphics[scale=0.9]{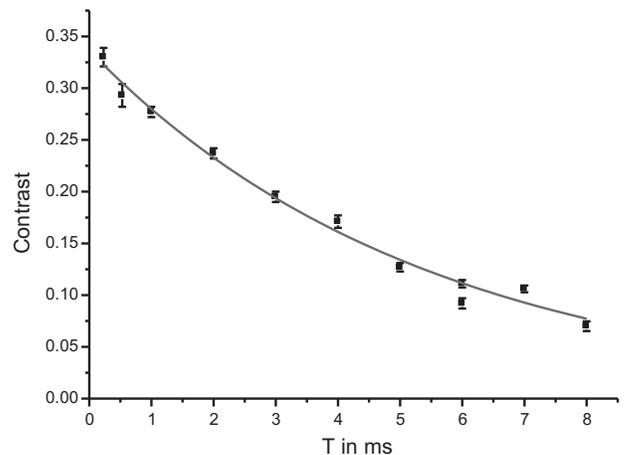}
 \caption{\label{contrastT} Contrast $C$ versus the time $T$ (75 Bloch oscillations, $T'=115\,\text{ms}-2T$).
 The solid line is an exponential fit with a decay constant of $\tau =5.4$ ms and an initial contrast value of 0.34.}
\end{figure}

The contrast of our interferometer has been measured versus the time T. The
result is plotted on figure \ref{contrastT}. The measurements are well fitted
with an exponential decay of 5.4 ms time constant. Thus, our experimental data
are consistent with a random velocity variation with a Lorentzian probability
distribution of $23\, \mu \text{m/s}$ FWHM. We will now examine what could be
the origin of this random velocity variation.

\subsection{Spontaneous emission}

The Bloch beams can interact with the atoms via the process of spontaneous
emission. The atoms absorb a photon from one of the lattice beams and emit a
spontaneous photon in a random direction. In this process, the velocity of the
atoms on the vertical axis can change randomly between $\pm 2 v_{rec}$. The
velocity width of the first band where the atoms are trapped is equal to
$2v_{rec}$. Thus, when a spontaneous emission occurs, approximately half of the
atoms will remain trapped in the lattice while the others will be lost. The
atoms remaining trapped in the lattice will have a random velocity variation
with a probability distribution with a width of approximately $v_{rec}$.
Finally, in presence of spontaneous emission, one can show that the contrast
will be given by:
\begin{eqnarray}
\frac{C}{C_0}&=&\frac{1}{1+\zeta (e^{\Gamma_{spt}N\,T_B}-1)}\\
&&+\frac{\zeta (e^{\Gamma_{spt}N\,T_B}-1)}{1+\zeta (e^{\Gamma_{spt}N\,T_B}-1)}\int P_{spt}(\delta v)cos(k_{eff}\delta v T)\,d\delta v \nonumber
\end{eqnarray}
where $\Gamma_{spt}$ is the rate of spontaneous emission in the lattice,
$\zeta$ is the proportion of atoms which remain trapped in the lattice after a
spontaneous emission and $P_{spt}$ is the probability distribution of the
velocity jumps which occur after a spontaneous emission and when the atoms
remain trapped in the lattice.

For our experimental parameters ($U_0 \sim 2 E_{rec}$, $\Delta=2\pi\cdot 58\;
 \text{GHz}$), the rate of spontaneous emission is equal to $\Gamma_{spt}\sim
3\;\text{s}^{-1}$. Thus, the contrast drop is limited to $0.87 C_0$ for 75
Bloch oscillations. The contrast decay versus T has a width of approximately
$1/k_{eff}v_{rec}=11\;\mu \text{s}$. In our experiment, the level of contrast
drops and the width of the contrast decay versus T does not correspond to the
characteristic of spontaneous emission. We can conclude that the contrast decay
observed experimentally does not come from spontaneous emission.

\subsection{Effect of a parasite reflection}

Parasite reflections coming from optics on the path of the Bloch beams can
interfere with the main beams and thus give interference fringes. These fringes
produce a vertical gradient of intensity which leads to a dipolar force which
changes the vertical velocity of the atoms and thus gives a phase shift on the
interferometer depending on the position of the atoms in the interference
fringes. Thus, parasite reflections can lead to a drop of contrast.

In our experiment, the main reflection comes from the window of the vacuum
chamber which has a reflection coefficient of $\sim 1\%$ per face. This window
has an angle of 15 mrad compared to the Bloch beams. The fringes coming from
the interference of the Bloch beam and the parasite reflection have a
horizontal period of $2\pi / k_x \sim 50\;\mu \text{m}$ and a vertical period
of $2\pi / k_z \sim 7\; \text{mm}$. The dipolar force exerted on the atoms by
the fringes is equal to :
\begin{equation}
F=\epsilon U_0 k_z sin(k_x x+ k_z z)
\label{force frange}
\end{equation}
where $\epsilon$ is the relative depth of the fringe which is in our case
approximately 7 \%.

The horizontal period of the fringes is small compared to the size of the
atomic cloud. Thus, the mean effect of the fringes will be canceled out and the
effect of the contrast drop will only remain. For 75 Bloch oscillations
(interaction time of 90 ms), one obtains with the force given in the equation
(\ref{force frange}) a maximum velocity shift of $196\; \mu \text{m/s}$. This
effect could thus be responsible of the random velocity variation of $23\;
\mu\text{m/s}$ deduced from the measurement of the contrast versus T.

We can conclude that fringes on the Bloch beams could explain the contrast
decay observed experimentally.

\subsection{Speckle}

The laser beam can also be affected by a speckle pattern coming from scattering
on optical elements. The speckle pattern is responsible of a dipolar force on
the atoms depending on their position in the speckle pattern. Thus, identically
to the interference fringes, the speckle can cause a contrast decay.

We will now give the order of magnitude of this effect in our experiment.
Scattering from the window of the vacuum chamber which is at a distance of 10
cm from the atoms gives a speckle pattern on the Bloch beam with a horizontal
correlation distance of $L_x \sim 9\; \mu\text{m}$ and a vertical correlation
distance of $L_z \sim 0.8\;\text{mm}$. The order of magnitude of the force from
the speckle is given by :
\begin{equation}
F\sim\frac{\epsilon .U_0}{L_z}
\end{equation}
where $\epsilon$ is the relative amplitude of the speckle pattern. For an
amplitude of the speckle pattern of $\epsilon=1\%$ and 75 Bloch oscillations,
one obtains a velocity variation of $39\; \mu\text{m/s}$. The order of
magnitude of this effect corresponds to the velocity variation of $23\; \mu
\text{m/s}$ deduced from the curve contrast versus T.

The effect of contrast decay with speckle has been confirmed experimentally by
the following observations. The contrast has increased by a factor of two for
75 Bloch oscillations and T=6 ms after a careful cleaning of the window of the
vacuum chamber. The insertion of a dusty window in the Bloch beam makes the
contrast to drop to a level where the fringes can not be observed.

\section{Conclusion}

In summary, we have presented an experimental demonstration of an atomic
gravimeter combining atom interferometry and Bloch oscillations. We have
developed a simple experimental setup where one retro-reflected laser beam
coming from a unique laser source is performing atom launch, stimulated Raman
transition and Bloch oscillations. We reach a sensitivity of $2\times 10^{-7}$g
in 300 seconds within a falling distance of 0.8 mm. We did not measure any bias
coming from the insertion of Bloch oscillations in the interferometer at our
level of sensitivity. Our error budget suggests that the insertion of Bloch
oscillations does not prevent from reaching $10^{-9}$g accuracy level with some
improvements on the experimental set-up. Our sensitivity is limited by a
contrast decay of the interference fringes coming from the insertion of Bloch
oscillations. This contrast decay can be attributed to imperfections of the
Bloch beams (fringes and speckle).

With a better beam quality obtainable with tilted windows and better optics,
this technique seems very promising to perform high precision local gravity
measurements. This technique should allow to make local force measurements with
high precision in order for example to test gravity at small distance \cite{PRD
gravity}. This technique should also allow to miniaturize atomic gravimeters.

We thank F. Nez,  P. Clad\'{e}, S. Guellati-Kh\'{e}lifa and F. Biraben for all
fruitful discussions and advices. This work was supported by DGA.


\begin{thebibliography}{}

\bibitem{best sens muller} H. M\"{u}ller, S. W. Chiow, S. Herrmann, S. Chu
    and K. Y. Chung, Phys. Rev Lett. \textbf{100}, 031101 (2008).

\bibitem{limit sens gravi syrte} J. Le Gou\"{e}t, T.E. Mehlst\"{a}ubler, J.
    Kim, S. Merlet, A. Clairon, A. Landragin and F. Pereira Dos Santos, Appl.
    Phys. B \textbf{92}, 133-144 (2008).

\bibitem{gravi peter} A. Peters, K.Y. Chung and S. Chu, Metrologia \textbf{38},
    25-61 (2001).

\bibitem{comp gravi syrte} S. Merlet, Q. Bodart, N. Malossi, A. Landragin, F.
    Pereira Dos Santos, O Gitlein and L Timmen, Metrologia \textbf{47}, L9-L11
    (2010).

\bibitem{mesure g biraben} P. Clad\'{e}, S. Guellati-Kh\'{e}lifa, C. Schwob,
    F. Nez, L. Julien and F. Biraben, Europhys. Lett. \textbf{71}, 730–736
    (2005).

\bibitem{mesure g tino} N. Poli, F.-Y. Wang, M. G. Tarallo, A. Alberti, M.
    Prevedelli, and G. M. Tino, Phys. Rev. Lett. \textbf{106}, 038501 (2011).


\bibitem{Bloch splitter biraben} P. Clad\'{e}, S. Guellati-Kh\'{e}lifa,
    F. Nez, and F. Biraben, Phys. Rev. Lett. \textbf{102}, 240402 (2009).

\bibitem{Bloch splitter chu} H. M\"{u}ller, S. W. Chiow, S.
    Herrmann, and S. Chu, Phys. Rev. Lett. \textbf{102}, 240403 (2009).

\bibitem{theorie latt interf} T. Kovachy, J. M. Hogan, D. M. S.
    Johnson, and M. A. Kasevich, Phys. Rev. A \textbf{82}, 013638 (2010).

\bibitem{det st fine biraben}M. Cadoret, E. de Mirandes, P. Clad\'{e}, S.
   Guellati-Kh\'{e}lifa, C. Schwob, F. Nez, L. Julien, F. Biraben, Phys. Rev.
   Lett. \textbf{101}, 230801 (2008); R. Bouchendira, P. Clad\'{e}, S. Guellati-Kh\'{e}lifa,
    F. Nez, and F. Biraben, Phys. Rev. Lett. \textbf{106}, 080801 (2011).



\bibitem{ramsey borde} Ch. J. Bord\'{e}, Phys. Lett. A \textbf{140}, 10 (1989).

\bibitem{Bloch oscillation nagerl} M. Gustavsson, E. Haller, M.J. Mark, J.G.
    Danzl, G. Rojas-Kopeinig, and H.-C. N\"{a}gerl, Phys. Rev. Lett.
    \textbf{100}, 080404 (2008).

\bibitem{laser onera} O. Carraz, F. Lienhart, R. Charri\`{e}re, M. Cadoret, N.
    Zahzam, Y. Bidel and A. Bresson, Appl. Phys. B \textbf{97}, 405-411 (2009).

\bibitem{Bloch Salomon PRA} E. Peik, M. Ben Dahan, I. Bouchoule, Y. Castin, C.
    Salomon, Phys. Rev. A \textbf{55}, 2989-3001 (1997).

\bibitem{det kasevich} G.W. Biedermann, X. Wu, L. Deslauriers, K. Takase, and
    M. A. Kasevich, Opt. Lett. \textbf{34}, 347-349 (2009).

 \bibitem{PRA h sur m} P. Clad\'{e}, E. de Mirandes, M. Cadoret, S. Guellati-Kh\'{e}lifa, C. Schwob,
    F. Nez, L. Julien and F. Biraben, Phys. Rev. A \textbf{74}, 052109
    (2006).

 \bibitem{PRD gravity} C. D. Hoyle, D. J. Kapner, B. R. Heckel, E. G. Adelberger, J. H. Gundlach, U. Schmidt,
  and H. E. Swanson, Phys. Rev. D \textbf{70}, 042004 (2004).

















\end{thebibliography}
\end{document}